\documentstyle[preprint,aps]{revtex}

\newcommand{\be}{\begin{equation}}
\newcommand{\ee}{\end{equation}}
\newcommand{\bea}{\begin{eqnarray}}
\newcommand{\eea}{\end{eqnarray}}

\begin{document}
 \preprint{\small CGPG-95/10-3  gr-qc/9511011}
\draft
\title
 {Exact solutions for null fluid collapse}
\author{Viqar Husain}
\address
 {Department of Mathematics and Statistics,\\
University of Calgary, Calgary, Alberta, Canada T2N 1N4,\\
 and\\
 Center for Gravitational Physics and Geometry,\\
The Pennsylvania State University, \\
 University Park, PA 16802-6300, USA.\footnote{Present address.}}
\maketitle
\begin{abstract}
 Exact non-static spherically symmetric solutions of the Einstein
equations for a null fluid source with pressure $P$ and
density $\rho$ related by $P = k\rho^a$  are given.
The $a=1$  metrics are asymptotically flat for $1/2<k\le 1$ and
cosmological for $0<k<1/2$.  The $k=1$ metric
is the known charged Vaidya solution.  In general the metrics
have multiple apparent horizons. In the long time limit, the
asymptotically flat metrics are hairy black hole solutions that
`fall between' the Schwarzschild and Reissner-Nordstrom metrics.
\end{abstract}
\bigskip

The problem of gravitational collapse in general relativity is of much
interest. One would like to know whether, and under what initial
conditions, gravitational collapse results in black hole formation. In
particular, one would like to know if there are physical collapse
solutions that lead to naked singularities. If found, such solutions
would be counter examples of the cosmic censorship hypothesis, which
states that curvature singularities in asymptotically flat spacetimes
are always shrouded by event horizons.

Since the general problem appears intractable due to the complexity of
the full Einstein equations, metrics with special symmetries are used
to construct gravitational collapse models. One such case is the
two-dimensional reduction of general relativity obtained by imposing
spherical symmetry. Even with this reduction however, there are very
few inhomogeneous non-static exact solutions known. One example is the
Vaidya metric \cite{vaidya,si}. It describes the collapse of
pressureless null dust and is asympotically flat. For the minimally
coupled massless scalar field, the only inhomogeneous non-static
examples are cosmological solutions which have naked singularities
\cite{roberts,vh}.  For perfect fluids with equation of state relating
pressure $P$ and energy density $\rho$ given by $P=k\rho$,
self-similar solutions (which are not asymptotically flat) have been
studied \cite{op}.

Recently there have been a number of numerical studies of the
spherically symmetric collapse problem \cite{chop,ae,ce,gar,eard}.
The initial data in these works is a compact ingoing pulse.  It was
found that a black hole forms if any of the parameters $c$ in the
initial data are above certain critical values $c_*$, otherwise the
pulse scatters back out to infinity leaving flat space.  In
particular, the black hole mass formula $M_{BH} = k(c-c_*)^{.36}$ was
discovered in Ref. \cite{chop}. This shows that black holes initially
form with zero mass - that is, no mass gap is found.

There have been a number of attempts aimed at an analytical
understanding these results
\cite{oshetal,pat,trach,pull,koimish,koike,maison,vhetal}. One
approach is to look for exact collapse solutions and see what black
hole mass formulas may be extracted from them.

In this paper we give exact inhomogeneous and non-static spherically
symmetric solutions of the Einstein equations for a collapsing null
fluid. The equation of state of the fluid is $P=k\rho^a$. As we will
see, while these collapse solutions do not have direct relevance for
the critical behaviour mentioned above, they do have a number of
interesting features, including hair on black holes.

An inverted approach is used to find the solutions. First the
stress-energy tensor is determined from the metric. Then the equation
of state and the dominant energy condition are {\it imposed} on its
eigenvalues. This leads to an equation for the metric function, which
is easily solved. The precise form of the stress-energy tensor is then
displayed. Some interesting properties of the solutions are then
discussed.

For the general spherically symmetric metric
\be
ds^2  =  -e^{2\psi(r,v)}F(r,v)\ dv^2 + 2 e^{\psi(r,v)}\ dvdr +
r^2 d\Omega^2, \label{metric}
\ee
where $0\le r\le \infty$ is the proper radial coordinate,
$-\infty\le v\le \infty$ is an advanced time
coordinate, and $d\Omega^2$ is the metric on the unit two sphere,
the Einstein equations $G_{ab} = 8\pi T_{ab}$ give
\bea
m'\equiv {\partial m\over \partial r } &=& -4\pi r^2T_v^{\ v},
\label{mp}\\
 \dot{m}\equiv {\partial m\over \partial v } &=& 4\pi r^2T_v^{\ r},
\label{md}\\
  \psi' &=& 4\pi r T_{rr}.\label{psip}
\eea

The mass function  $m(r,v)$ is  defined by $F(r,v) = 1-2m(r,v)/r$,
and is measure of the mass contained within radius $r$.  We will
consider the special case $\psi(r,v) = 0$, which means from
(\ref{psip}) that $T_{rr}=0$.

The stress-energy tensor derived from the above metric may be
diagonalized to give the energy density and the principal
pressures. The eigenvalue problem is $T_a^{\ b}U_b = \lambda U_a$.
The $\theta$-$\phi$ part of $T_a^{\ b}$, which is determined from
$G_{\theta\theta}$ and $G_{\phi\phi}=\sin^2\theta\ G_{\theta\theta}$,
is already diagonal with pressure eigenvalues
\be
 P\equiv T_\theta^{\ \theta} = T_\phi^{\ \phi} = -{m''\over 8\pi r}.
\label{P}
\ee
Since $T_{rr}=0$ (by choice), we have $T_v^{\ v} =  T_r^{\ r}$ and
$T_r^{\ v}=0$. Therefore the $v$-$r$ part of the matrix to be diagonalized
is
\be
T_a^{\ b} = \left( \begin{array}{cc}
                    T_v^{\ v} & T_v^{\ r}   \\
                    0 &  T_v^{\ v}
                    \end{array}
             \right),
\ee
This has one eigenvalue $\lambda$ which gives the energy density
$\rho$, namely
\be
 -\rho \equiv \lambda = T_v^{\ v} = -{m'\over 4\pi r^2}\ .
\label{rho}
\ee
The corresponding  eigenvector is $v_a=(1,0,0,0)$ (in the
coordinates $(v,r,\theta,\phi)$), and is lightlike. Therefore
the  stress-energy  tensor, (which follows from $\psi=0$), is
of Type II \cite{hawkell}. Its non-vanishing components are
\bea
T_{vv} &=& \rho\ (1- {2m\over r}\ ) + {\dot{m}\over 4\pi r^2},
\ \ \ \ \ \ T_{vr} = -\rho,
\\ \nonumber
T_{\theta\theta} &=& P\ g_{\theta\theta}, \ \ \ \ \ \
T_{\phi\phi} = P\ g_{\phi\phi},
\eea
These components  may  be succinctly written using the two linearly
independent future pointing lightlike vectors $v_a=(1,0,0,0)$ and
$w_a=(F/2,-1,0,0)$ as
\be
T_{ab} =  {\dot{m}(r,v)\over 4\pi r^2} \ v_av_b +
\rho(r,v)\ (v_aw_b + v_bw_a) + P(r,v)\ (g_{ab} + v_aw_b + v_bw_a)\ .
\label{set}
\ee
(This  tensor should be compared with the perfect fluid
one $T_{ab} = \rho\ u_au_b + P\ (\ g_{ab} + u_au_b\ )$,
 where $u_a$ is timelike.)

The stress-energy tensor (\ref{set}) has support along both the two
future pointing lightlike vectors $v_a$ and $w_a$, and as we will see
below, it is of precisely the form which gives the charged Vaidya
solution \cite{si}. For $P=\rho=0$, (\ref{set}) reduces to the
stress-energy tensor which gives the uncharged Vaidya metric.  We also
note that while $T_{ab}w^aw^b=\dot{m}/4\pi r^2$, $T_{ab}v^av^b=0$.
Therefore there is energy flux only along one of the null directions.

A static observer with 4-velocity $S^a=(1/\sqrt{F},0,0,0)$ and a
rotating observer with 4-velocity $R^a=(\sqrt{2/F},0,0,1/ r
\sin\theta)$ see respectively the energy densities $T_{ab}S^aS^b=
\dot{m}/4\pi Fr^2 + \rho$ and, $T_{ab}R^aR^b = 2(\dot{m}/4\pi Fr^2 +
\rho ) + P$

The stress-energy tensor (\ref{set}) satisfies the dominant energy
condition if the following three conditions are met :
\be
  P\ge 0, \ \ \rho\ge P, \ \ {\rm and} \ \ T_{ab}w^a w^b > 0,
\label{de}
\ee
 The first two of these imply that $m'\ge 0$
and $m''\le 0$. The former just says that the mass function either
increases with $r$ or is constant,  which is a natural physical
requirement on it.

To  satisfy the first two of the dominant energy conditions
(\ref{de}), we {\it impose} the  equation of state $P=k\rho^a$, with
$k\le 1$ and $a\le 1$ .  The $a<1$ case will be discussed later.
For $a=1$ this gives the equation
\be
-{m''\over 8\pi r} = k\ { m'\over 4\pi r^2}.
\ee
for the mass function, which is easily integrated to give
\be
m(r,v) = \left\} \begin{array}{cc}
 f(v)-g(v)/ [\ (2k-1)\ r^{2k-1}\ ]\ , &   \ k\ne {1\over 2} \\
  f(v) + g(v) \ln r \ ,&    \ k = {1\over 2}
                \end{array} \right.
\label{mfn}
\ee
 where $f(v)$ and $g(v)$ are arbitrary functions (which are restricted
only by the energy conditions). Therefore we have explicitly that
\be
P = k\ {g(v)\over 4\pi r^{2k+2} } = k \rho.
\label{Prho}
\ee
Therefore we must have $g(v)\ge 0$ for positive pressure and
energy density.

The last requirement in (\ref{de}) for the dominant energy
condition leads, for $k\ne 1/2$ to
\be
\dot{m} = \dot{f}(v) - {\dot{g}(v) \over (2k-1)\ r^{2k-1} } > 0.
\ee
Physically this means that the matter within a radius $r$ increases
with time, which corresponds to an implosion.  This condition is most
easily satisfied if $\dot{f}> 0$, and either $\dot{g}> 0$ and $k<
1/2$, or $\dot{g}< 0$ and $k > 1/2$.

In summary, for the dominant energy condition, we must have $g(v)\ge
0$ and, either $\dot{g}>0$ for $k<1/2$ or $\dot{g}< 0$ for
$k>1/2$. For the weak or strong energy conditions, (which are
equivalent for Type II stress-energy tensors), we only need $\rho\ge
0$, $P\ge 0$ and $T_{ab}w^aw^b>0$, but not $\rho >P$. Therefore for
the latter energy conditions we can have $k>1$ as well.

For $k=1/2$, neither the weak nor dominant energy conditions can be
satisfied for all $r$ because $\dot{m} = \dot{f}(v)+ \dot{g}(v)\ln r$,
which always becomes negative for sufficiently small $r$. Therefore
we will not consider this case further.

In summary, we have shown that the metric
\be
ds^2 = -(\ 1-{2f(v)\over r} + {2g(v)\over (2k-1)\ r^{2k}}\ )\ dv^2
+ 2\ dv dr + r^2 d\Omega^2 \label{sol1}
\ee
 is a solution of the Einstein equations for the null fluid
 stress-energy tensor (\ref{set}) with $P=k\rho$, where $P$ is given
by (\ref{Prho}).

There are two special cases of this solution which are already known.
One is the Vaidya metric \cite{vaidya}, which arises for $g(v)=0$
(vanishing $\rho$ and $P$). Then the only non-vanishing component of
the stress-energy tensor (\ref{set}) is $T_{vv} = \dot{m}/4\pi
r^2=\dot{f}(v)/4\pi r^2$. The other is the charged Vaidya metric,
where the charge depends on $v$ \cite{si}.  This arises when $k=1$ in
(\ref{sol1}).

We therefore see that the stress-energy tensor we have determined by
imposing the equation of state $P=k\rho$ is a one parameter $(k>0)$
generalization of the stress-energy tensor which gives the charged
Vaidya metric. The corresponding metric must therefore depend on $k$
as in equation (\ref{sol1}). (We note that the parameter $k$ in the metric
or the stress-energy tensor cannot be eliminated by a coordinate
transformation because it is the constant of proportionality between
the eigenvalues of the stress-energy tensor - exactly the same
reason that this constant isn't `gauge' for the ordinary perfect
fluid solutions with equation of state $P=k\rho$.)

A metric is considered to be asymptotically flat \cite{beig,witt} if
in the vicinity of a spacelike hypersurface its components behave as
\be
g_{ab} \rightarrow \eta_{ab} + {\alpha_{ab}(x^c/r, t)\over r}
+ {\cal O}({1\over r^{1+\epsilon}}) \nonumber
\ee
as $r\rightarrow\infty$. ($\epsilon>0$, $\eta_{ab}$ is the
 Minkowski metric, $\alpha_{ab}$ is an arbitrary symmetric tensor, and
$x^c$ is a flat coordinate system at spacelike infinity.)

According to this definition, our metrics (\ref{sol1}) are
asymptotically flat for $k>1/2$ and are cosmological for $k<1/2$.
(In particular, for $k=1$, $f(v)=M$ and $2g(v) = Q^2$, the metric
is just Reissner-Nordstrom).

When the imposed equation of state on the eigenvalues
(\ref{P}) and (\ref{rho}) of the stress-energy tensor (\ref{set})
is  $P=k\rho^a$,  the  equation for the mass function is more
complicated :
\be
-{m''\over 8\pi r} = k\ \bigl( {m' \over 4\pi r^2} \bigr)^a.
\ee
 This has solution
\be
m(r,v) = f(v) + \int dr
\bigl[\ g(v) - k\ (4\pi)^{1-a} r^{2(1-a)}  \  \bigr]^{1/1-a}.
\label{mass2}
\ee
Since $a<1$, this mass function  gives only cosmological
metrics - the pressures are too small to make them
asymptotically flat. However if the dominant energy condition
is not imposed but only the weak (or strong) one is, then $a>1$
is possible, which will give asymptotically flat metrics.

We now consider in turn the asymptotically flat $(k>1/2)$ and
the cosmological $(k< 1/2)$ metrics.

\noindent \underbar{$k>1/2$}: Since the dominant energy condition is
satisfied only for $g(v) \ge 0$ and $\dot{g}(v)\le 0$, if $g(v)$ is zero
 initially, it must remain zero. On the other hand, if it is non-zero
initially, it can decrease to zero. Therefore the asymptotically flat
metrics can be flat in the $v\rightarrow - \infty$ limit only if
$g(v)\equiv 0$. But this is just the Vaidya case.  However the
following two parameter $(A \ge 0, 0\le B\le 1)$ family of metrics,
(where we have set $k=1$ for concreteness), demonstrates an
interesting feature:
\be
ds^2 = -(1-{A(1+ {\rm tanh}v)\over r} +
{ (1-B\ {\rm tanh}v)\over r^2} )\ dv^2 + 2\ dv dr + r^2 d\Omega^2
\label{k=1}
\ee
In the $v\rightarrow -\infty$ limit this metric has a naked
singularity at $r=0$. However, in the $v\rightarrow +\infty$ limit
it may have horizons depending on the relative values of $A,B$.
Specifically, these horizons are given by $ r = A \pm \sqrt{A^2 + B-1}$.
 Therefore a black hole first forms at non-zero mass when $A^2=1-B$.
This mass gap seperates a black hole from a naked singularity, and
 may be contrasted  with the critical behaviour solutions
\cite{chop,ae,ce} where there is no mass gap between flat space
and a black hole metric. Similar results hold for other
values of $k>1/2$. What is happening physically is that initially
only the `charge' term is present in the metric, and subsequently,
infalling fluid reduces the `charge' and adds a `mass' term.

These solutions also include metrics that give the evolution of a
naked singularity at $v=-\infty$ into flat space at $v=\infty$. This
case occurs if we set $A=0, B=1$ in (\ref{k=1}).

Another feature of these solutions is that they give black holes
with null fluid hair for $1/2<k<1$. An example of such a static solution
results from the $v\rightarrow\infty$ limit of the metric
\be
ds^2 = -(1-{A(1+ {\rm tanh}v)\over r} + {(1-B\ {\rm tanh}v)\over r^{3/2} })
\ dv^2 + 2\ dv dr + r^2 d\Omega^2,
\ee
 with $B$ as the `hair'. (We have put $k=3/4$).

\noindent \underbar{$k<1/2$}: The dominant energy condition is now
satisfied for $g(v)\ge 0$ and $\dot{g}(v)\ge 0$. Therefore these
metrics can be flat as $v\rightarrow -\infty$, and have black holes
as $v\rightarrow\infty$. A specific two parameter $(A,B\ge 0$) example
for $k=1/3$ (electromagnetic radiation) is
\be
ds^2 = -(\ 1 - {C+ A(1 + {\rm tanh} v)\over r} -
{B(1 + {\rm tanh} v)\over r^{2/3}} \ )\ dv^2
+ 2\ dvdr +  r^2 d\Omega^2.
\label{k=1/3}
\ee
The apparent horizons in the $v\rightarrow\infty$ limit (with $C=0$)
are now given by the cubic equation $(r-2A)^3 -(2B)^3 r = 0$. This
equation always has a solution for the ranges of $A$ and $B$ allowed by
the energy conditions. Thus, this solution describes the collapse of
electromagnetic radiation  ($P=\rho/3$) from flat space at $v=-\infty$
to a black hole at $v=\infty$.

Another possibility occurs when $C$ is negative in (\ref{k=1/3}).
Then, as for the $k>1/2$ case above, we have a naked singularity at
$v=-\infty$, which becomes a black hole at $v=\infty$ as the
collapse proceeds.

In the general case, if $\lim_{v\rightarrow\infty} f(v)=A$ and
$\lim_{v\rightarrow\infty} g(v)=B$ the radii of the apparent horizons
in the $v\rightarrow \infty$ limit of the metric (\ref{sol1}) are
given  by
\be
r = 2A - {2B\over (2k - 1)\ r^{2k-1} }\ , \label{ah1}
\ee
which in general may have multiple solutions.

In conclusion, we have given a new class  of null fluid
collapse solutions (\ref{sol1}) and (\ref{mass2}) of the Einstein
equations for the stress-energy tensor (\ref{set}). These include
new asymptotically flat black hole solutions ($a=1, 1/2< k < 1$)
with multiple apparent horizons and hair. The general metric depends
on one parameter ($k$), and  two arbitrary functions of $v$
(modulo energy conditions). The long time limits of the asymptotically
flat solutions `fall between' the Schwarzschild and Reissner-Nordstrom
metrics in the sense that  $ 1< 2k \le 2 $ in (\ref{sol1}).

Physically, the $k>1/2$ solutions describe the evolution of a naked
singularity into the same, or into a black hole.  The parameters in
the solution (\ref{k=1}) determine which of these possibilities
occurs, and the black hole always forms at a finite non-zero mass. The
$k<1/2$ solutions describe the evolution of either flat space or a
naked singularity into a black hole in a cosmology. All of the new
solutions support the cosmic censorship conjecture.

Because of these physical properties, the exact solutions we have
given are not of relevance for the collapse solutions that exibit
critical behaviour \cite{chop,ae,ce}. The physical reason for this
is that, although our stress-energy tensor (\ref{set}) has non-zero
components along both ingoing and outgoing null directions, there is
energy flow only along one direction, (because, as noted above
 $T_{ab}v^av^b=0$).

It should  be possible to find more exact solutions by imposing
other equations of state on the eigenvalues of the stress-energy
tensor (\ref{set}).

\bigskip

This work was supported by the Natural Science and Engineering
Research Council of Canada, and by NSF grant PHY 93-96246 to the
Pennsylvania State University.

\end{document}